\documentclass[aps,prl,showpacs,showkeys,amsmath,amssymb,
twocolumn,
floatfix,
superscriptaddress
]{revtex4-1}
\usepackage{graphicx}
\usepackage{times}
\usepackage{verbatim}
\usepackage{color}
\usepackage{cancel}
\usepackage[normalem]{ulem}

\newcommand{\nuebar}{$\overline{\nu}_{e}$}
\newcommand{\myol}[2][3]{{}\mkern#1mu\overline{\mkern-#1mu#2}} 

\begin{document}

\title{A new measurement of antineutrino oscillation with the full detector configuration at Daya Bay}

\newcommand{\ECUST}{\affiliation{Institute of Modern Physics, East China University of Science and Technology, Shanghai}}
\newcommand{\IHEP}{\affiliation{Institute~of~High~Energy~Physics, Beijing}}
\newcommand{\Wisconsin}{\affiliation{University~of~Wisconsin, Madison, Wisconsin, USA}}
\newcommand{\Yale}{\affiliation{Department~of~Physics, Yale~University, New~Haven, Connecticut, USA}}
\newcommand{\BNL}{\affiliation{Brookhaven~National~Laboratory, Upton, New York, USA}}
\newcommand{\NTU}{\affiliation{Department of Physics, National~Taiwan~University, Taipei}}
\newcommand{\NUU}{\affiliation{National~United~University, Miao-Li}}
\newcommand{\Dubna}{\affiliation{Joint~Institute~for~Nuclear~Research, Dubna, Moscow~Region}}
\newcommand{\CalTech}{\affiliation{California~Institute~of~Technology, Pasadena, California, USA}}
\newcommand{\CUHK}{\affiliation{Chinese~University~of~Hong~Kong, Hong~Kong}}
\newcommand{\NCTU}{\affiliation{Institute~of~Physics, National~Chiao-Tung~University, Hsinchu}}
\newcommand{\NJU}{\affiliation{Nanjing~University, Nanjing}}
\newcommand{\TsingHua}{\affiliation{Department~of~Engineering~Physics, Tsinghua~University, Beijing}}
\newcommand{\SZU}{\affiliation{Shenzhen~University, Shenzhen}}
\newcommand{\NCEPU}{\affiliation{North~China~Electric~Power~University, Beijing}}
\newcommand{\Siena}{\affiliation{Siena~College, Loudonville, New York, USA}}
\newcommand{\IIT}{\affiliation{Department of Physics, Illinois~Institute~of~Technology, Chicago, Illinois, USA}}
\newcommand{\LBNL}{\affiliation{Lawrence~Berkeley~National~Laboratory, Berkeley, California, USA}}
\newcommand{\UIUC}{\affiliation{Department of Physics, University~of~Illinois~at~Urbana-Champaign, Urbana, Illinois, USA}}
\newcommand{\RPI}{\affiliation{Department~of~Physics, Applied~Physics, and~Astronomy, Rensselaer~Polytechnic~Institute, Troy, New~York, USA}}
\newcommand{\SJTU}{\affiliation{Shanghai~Jiao~Tong~University, Shanghai}}
\newcommand{\BNU}{\affiliation{Beijing~Normal~University, Beijing}}
\newcommand{\WM}{\affiliation{College~of~William~and~Mary, Williamsburg, Virginia, USA}}
\newcommand{\Princeton}{\affiliation{Joseph Henry Laboratories, Princeton~University, Princeton, New~Jersey, USA}}
\newcommand{\VirginiaTech}{\affiliation{Center for Neutrino Physics, Virginia~Tech, Blacksburg, Virginia, USA}}
\newcommand{\CIAE}{\affiliation{China~Institute~of~Atomic~Energy, Beijing}}
\newcommand{\SDU}{\affiliation{Shandong~University, Jinan}}
\newcommand{\NanKai}{\affiliation{School of Physics, Nankai~University, Tianjin}}
\newcommand{\UC}{\affiliation{Department of Physics, University~of~Cincinnati, Cincinnati, Ohio, USA}}
\newcommand{\DGUT}{\affiliation{Dongguan~University~of~Technology, Dongguan}}
\newcommand{\XJTU}{\affiliation{Xi'an Jiaotong University, Xi'an}}
\newcommand{\UCB}{\affiliation{Department of Physics, University~of~California, Berkeley, California, USA}}
\newcommand{\HKU}{\affiliation{Department of Physics, The~University~of~Hong~Kong, Pokfulam, Hong~Kong}}
\newcommand{\UH}{\affiliation{Department of Physics, University~of~Houston, Houston, Texas, USA}}
\newcommand{\Charles}{\affiliation{Charles~University, Faculty~of~Mathematics~and~Physics, Prague}}
\newcommand{\USTC}{\affiliation{University~of~Science~and~Technology~of~China, Hefei}}
\newcommand{\TempleUniversity}{\affiliation{Department~of~Physics, College~of~Science~and~Technology, Temple~University, Philadelphia, Pennsylvania, USA}}
\newcommand{\CUC}{\affiliation{Instituto de F\'isica, Pontificia Universidad Cat\'olica de Chile, Santiago, Chile}} 
\newcommand{\CGNPG}{\affiliation{China General Nuclear Power Group}}
\newcommand{\NUDT}{\affiliation{College of Electronic Science and Engineering, National University of Defense Technology, Changsha}} 
\newcommand{\IowaState}{\affiliation{Iowa~State~University, Ames, Iowa, USA}}
\newcommand{\ZSU}{\affiliation{Sun Yat-Sen (Zhongshan) University, Guangzhou}}
\newcommand{\CQU}{\affiliation{Chongqing University, Chongqing}} 
\author{F.~P.~An}\ECUST
\author{A.~B.~Balantekin}\Wisconsin
\author{H.~R.~Band}\Yale
\author{M.~Bishai}\BNL
\author{S.~Blyth}\NTU\NUU
\author{I.~Butorov}\Dubna
\author{G.~F.~Cao}\IHEP
\author{J.~Cao}\IHEP
\author{W.~R.~Cen}\IHEP
\author{Y.~L.~Chan}\CUHK
% ...

\collaboration{The Daya Bay Collaboration}\noaffiliation
\date{\today}

\begin{abstract}
\noindent We report a new measurement of electron antineutrino
disappearance using the fully-constructed Daya Bay Reactor Neutrino
Experiment.  The final two of eight antineutrino detectors were
installed in the summer of 2012. Including the 404 days of data
collected from October~2012 to November~2013 resulted in a total
exposure of 6.9$\times$10$^5$ GW$_{\rm th}$-ton-days, a 3.6 times
increase over our previous results.  Improvements in energy
calibration limited variations between detectors to 0.2\%.  Removal of
six $^{241}$Am-$^{13}$C radioactive calibration sources reduced the
background by a factor of two for the detectors in the experimental
hall furthest from the reactors.  Direct prediction of the
antineutrino signal in the far detectors based on the measurements in
the near detectors explicitly minimized the dependence of the
measurement on models of reactor antineutrino emission.  The
uncertainties in our estimates of $\sin^{2}2\theta_{13}$ and $|\Delta
m^2_{ee}|$ were halved as a result of these improvements. Analysis of
the relative antineutrino rates and energy spectra between detectors
gave $\sin^{2}2\theta_{13} = 0.084\pm0.005$ and $|\Delta m^{2}_{ee}|=
(2.42\pm0.11) \times 10^{-3}\ {\rm eV}^2 $ in the three-neutrino
framework.
\end{abstract}

\maketitle

Neutrino flavor oscillation due to the mixing angle $\theta_{13}$ has
been observed using reactor antineutrinos~\cite{DB, RENO, Abe:2014bwa}
and accelerator neutrinos~\cite{Abe:2013hdq,Adamson:2013ue}.  The Daya
Bay experiment previously reported the discovery of a non-zero value
of $\sin^22\theta_{13}$ by observing the disappearance of reactor
antineutrinos over kilometer distances~\cite{DB,CPC,An:2014ehw}, and
the first measurement of the effective mass splitting
$|{\Delta}m^2_{ee}|$~\footnote{$\Delta m^2_{ee}$ is an effective mass splitting that can be obtained
  by replacing  $\cos^2\theta_{12}\sin^2
  \Delta_{31}+\sin^2\theta_{12}\sin^2{\Delta_{32}}$ with
$\sin^2\Delta_{ee}$, where
  $\Delta_{ji}\equiv 1.267 {\Delta}m^2_{ji}({\rm eV}^2) [L({\rm
      m})/E({\rm MeV})]$, and ${\Delta}m^2_{ji}$ is the difference
  between the mass-squares of the mass eigenstates $\nu_j$ and
  $\nu_i$.
  To estimate the values of ${\Delta}m^2_{31}$ and ${\Delta}m^2_{32}$
   from the measured value of ${\Delta}m^2_{ee}$,
   see the description in Appendix.}
via the distortion of the $\overline{\nu}_e$ energy
spectrum~\cite{DBPRL2014}.  Here we present new results with
significant improvements in energy calibration and background
reduction.  Installation of the final two detectors and a tripling of
operation time provided a total exposure of 6.9$\times$10$^5$ GW$_{\rm
  th}$-ton-days, 3.6 times more than reported in our previous
publication~\cite{DBPRL2014}.  With these improvements the precision
of $\sin^22\theta_{13}$ was enhanced by a factor of two compared to
the world's previous best estimate.  The precision of
$|{\Delta}m^2_{ee}|$ was equally enhanced, and is now competitive with
the precision of $|\Delta m^{2}_{32}|$ measured via accelerator
neutrino disappearance~\cite{Adamson:2014vgd, Abe:2014ugx}.

The Daya Bay experiment started collecting data on 24 December 2011
with six antineutrino detectors (ADs) located in three underground
experimental halls (EHs).  Three ADs were positioned in two near halls
at short distances from six nuclear reactor cores, two ADs in EH1 and
one in EH2, and three ADs were positioned in the far hall, EH3.  Data
taking was paused on 28 July 2012 while two new ADs were installed,
one in EH2 and the other in EH3. During the installation, a broad set
of calibration sources were deployed into the two ADs of EH1 using
automated calibration units~\cite{Liu:2013ava} and a manual
calibration system~\cite{Huang:2013uxa}.  Operation of the full
experiment with all eight ADs started on 19 October 2012. This Letter
presents results based on 404 days of data acquired in the 8-AD period
combined with all 217 days of data acquired in the 6-AD period.  A
blind analysis strategy was implemented by concealing the baselines
and target masses of the two new ADs, as well as the operational data
of all reactor cores for the new data period.

Each of the three Daya Bay experimental halls hosts functionally
identical ADs inside a muon detector system. The latter consists of a
two-zone pure water Cherenkov detector, referred to as the inner and
outer water shields (IWS and OWS), covered on top by an array of
resistive plate chambers~(RPCs).  Each AD consists of three nested
cylindrical vessels. The inner vessel is filled with 0.1\%
gadolinium-doped liquid scintillator (Gd-LS), which constitutes the
primary antineutrino target.  The vessel surrounding the target is
filled with undoped LS, increasing the efficiency of detecting gamma
rays produced in the target. The outermost vessel is filled with
mineral oil.  A total of 192 20-cm photomultiplier tubes (PMTs) are
radially positioned in the mineral-oil region of each AD.  Further
details on the experimental setup are contained in
Refs.~\cite{DBNIM,Dayabay:2014vka,dyb:det,DBproposal}.  Reactor
antineutrinos are detected via the inverse $\beta$-decay (IBD)
reaction, $\overline{\nu}_{e} + p \to e^{+} + n$. The gamma rays
(totalling $\sim\!8$~MeV) generated from the neutron capture on Gd
with a mean capture time of $\sim$30~$\mu{\rm s}$ form a delayed
signal and enable powerful background suppression. The light from the
$e^{+}$ gives an estimate of the incident \nuebar\ energy,
$E_{\overline{\nu}_e} \approx E_{\rm p} + \myol{E}_n + 0.78$~MeV,
where $E_{\rm p}$ is the prompt energy including the positron kinetic
and annihilation energy, and $\myol{E}_n$ is the average neutron
recoil energy ($\sim$10~keV).


\begin{figure}[htb]
\includegraphics[width=\columnwidth]{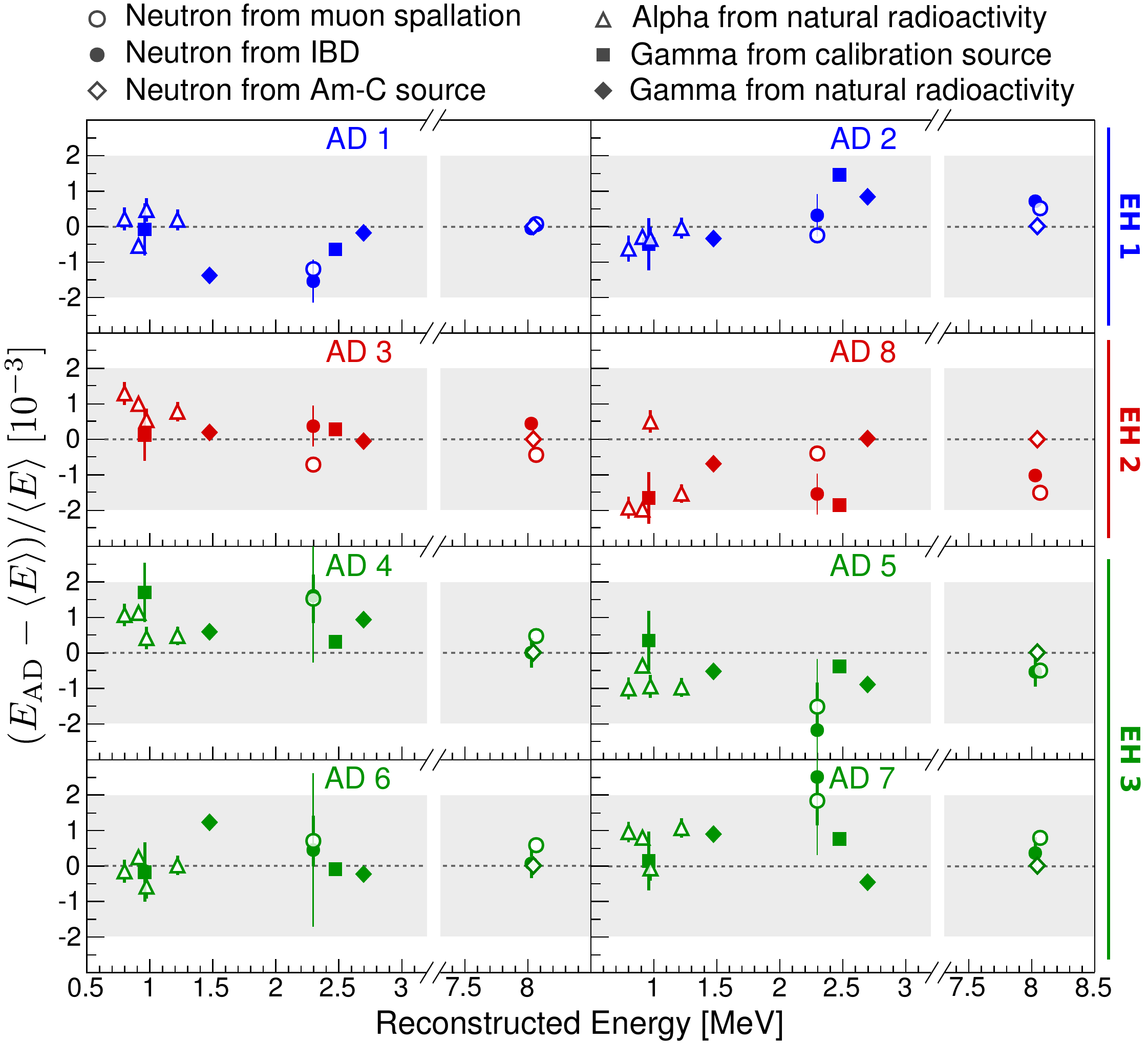}
\caption{ Comparison of the reconstructed energy between antineutrino
  detectors for a variety of calibration references.  $E_{\rm AD}$ is
  the reconstructed energy determined using each AD, and $\left\langle
  E \right\rangle$ is the 8-detector average.  Error bars are
  statistical only, and systematic variations between detectors for
  all calibration references were $<$~0.2\%.  The $\sim$8~MeV n-Gd
  capture gamma peaks from Am-C sources were used to define the energy
  scale of each detector, and hence show zero deviation.
  \label{fig:escale}}
\end{figure}

Differences in energy response between detectors directly impacted the
estimation of $|\Delta m^{2}_{ee}|$.  PMT gains were calibrated
continuously using uncorrelated single electrons emitted by the
photocathode. The signals of 0.3\% of the PMTs were discarded due to
abnormal hit rates or charge distributions. The detector energy scale
was calibrated using Am-C neutron sources~\cite{dyb:amc} deployed at
the detector center, with the $\sim$8~MeV peaks from neutrons captured
on Gd aligned across all eight detectors. The time variation and the
position dependence of the energy scale was corrected using the
2.506~MeV gamma-ray peak from $^{60}$Co calibration sources.  The
reconstructed energies of various calibration reference points in
different ADs are compared in Fig.~\ref{fig:escale}.  The spatial
distribution of each calibration reference varies, incorporating
deviations in spatial response between detectors.  Figure 1 presents measurements of $^{68}$Ge,
$^{60}$Co and Am-C calibration sources when placed at the center of
each detector. Neutrons from IBD and muon spallation that were
captured on gadolinium, were distributed nearly uniformly throughout the Gd-LS region.
Those neutrons that were captured on $^1$H,
intrinsic $\alpha$ particles from polonium and radon decays,
and gammas from $^{40}$K and $^{208}$Tl decays,
were distributed inside and outside of the target volume.
All of these events were selected within the Gd-LS region based on their reconstructed vertices.
The uncorrelated relative uncertainty of the energy scale
is thus determined to be $0.2\%$.  This reduction of 43\% compared to
the previous publication~\cite{DBPRL2014} was enabled by improvements
in the correction of position and time dependence, and enhanced the
precision of $|{\Delta}m^2_{ee}|$ by 9\%. The reduction was confirmed
by an alternative method which used the n-Gd capture of muon-induced
spallation neutrons to calibrate the scale, time dependence, and
spatial dependence of the detector energy response.

Nonlinearity in the energy response of an AD originated from two
dominant sources: particle-dependent nonlinear light yield of the
scintillator and charge-dependent nonlinearity in the PMT readout
electronics. Each effect was at the level of 10\%. We constructed a
semi-empirical model that predicted the reconstructed energy for a
particle assuming a specific energy deposited in the scintillator. The
model contained four parameters: Birks' constant, the relative
contribution to the total light yield from Cherenkov radiation, and
the amplitude and scale of an exponential correction describing the
non-linear electronics response.
This exponential form of the electronics response was motivated by MC
and confirmed with an independent FADC measurement.

\begin{figure}
\includegraphics[width=0.95\columnwidth]{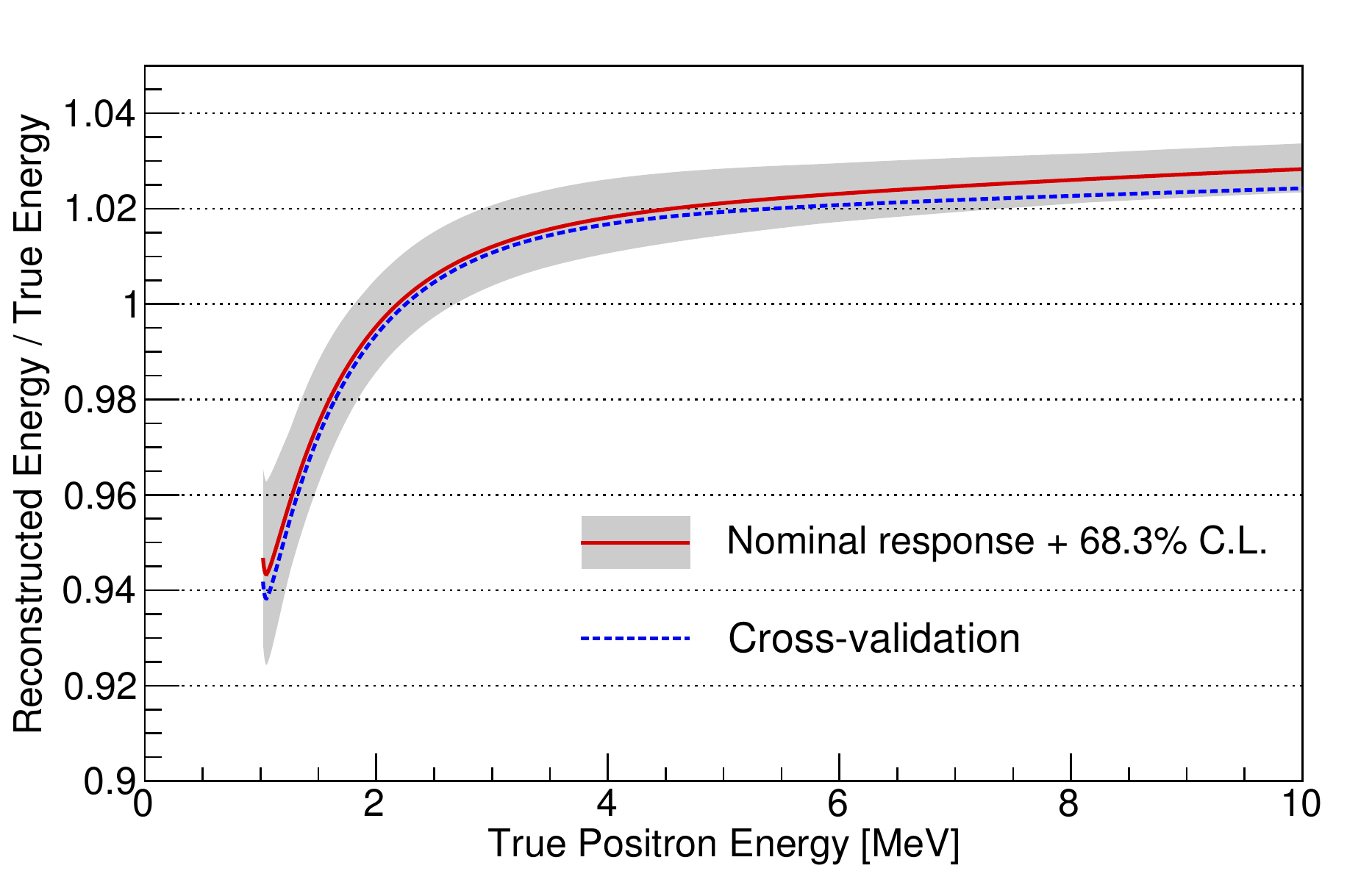}
\caption{ Estimated energy response of the detectors to positrons,
  including both kinetic and annihilation gamma energy (red solid
  curve).  The prominent nonlinearity below 4~MeV was attributed to
  scintillator light yield (from ionization quenching and Cherenkov
  light production) and the charge response of the electronics.  Gamma
  rays from both deployed and intrinsic sources as well as spallation
  $^{12}$B $\beta$ decay determined the model, and provided an
  envelope of curves consistent with the data within a 68.3\%
  C.L. (grey band).  An independent estimate using the beta+gamma
  energy spectra from $^{212}$Bi, $^{214}$Bi, $^{208}$Tl, as well as
  the 53-MeV edge in the Michel electron spectrum gave a similar
  result (blue dashed line), albeit with larger systematic
  uncertainties.\label{fig:ibdresponse}}
\end{figure}

The nominal parameter values were obtained from an unconstrained
$\chi^2$-fit to various AD calibration datasets, comprising twelve
gamma lines from both deployed and naturally occurring sources as well
as the continuous $\beta$-decay spectrum of $^{12}$B produced by muon
spallation inside the Gd-LS volumes. The nominal positron response
derived from the best fit parameters is shown in
Fig.~\ref{fig:ibdresponse}. The depicted uncertainty band represents
other response functions consistent with the fitted calibration data
within a 68.3\% C.L. This $\chi^2$-based approach to obtain the energy
response resulted in $<$~1\% uncertainties of the absolute energy
scale above 2~MeV.  The uncertainties of the positron response were
validated using the 53 MeV cutoff in the Michel electron spectrum from
muon decay at rest and the continuous $\beta$+$\gamma$ spectra from
natural bismuth and thallium decays.  These improvements added
confidence in the characterization of the absolute energy response of
the detectors, although they resulted in negligible changes to the
measured mixing parameters.

IBD candidates were selected using the same criteria discussed in
Ref.~\cite{DB}.  Noise introduced by PMT light emission in the voltage
divider, called {\it flashing}, was efficiently removed using the
techniques of Ref.~\cite{CPC}.  We required 0.7 MeV $<\!E_{\rm p}\!< $
12.0 MeV, 6.0 MeV $<\!E_{\rm d}\!<$ 12.0 MeV, and 1 $\mu$s $< \!
\Delta t \!<$ 200 $\mu$s, where $E_{\rm d}$ is the delayed energy and
$\Delta t = t_{\rm d} - t_{\rm p}$ was the time difference between the
prompt and delayed signals.  In order to suppress cosmogenic products,
candidates were rejected if their delayed signal occurred (i) within a
($-2~\mu{\rm s}$,~$600~\mu{\rm s}$) time-window with respect to an IWS
or OWS trigger with a PMT multiplicity $>$~12, (ii) within a
($-2~\mu{\rm s}$,~$1000~\mu{\rm s}$) time-window with respect to
triggers in the same AD with reconstructed energy $>$~20 MeV, or (iii)
within a ($-2~\mu{\rm s}$,~$1~{\rm s}$) time-window with respect to
triggers in the same AD with reconstructed energy $>$~2.5 GeV.  To
select only definite signal pairs, we required the signal to have a
{\em multiplicity} of 2: no other $>$~0.7 MeV signal occurred within a
($t_{\rm p} - 200~\mu s, t_{\rm d} + 200~\mu s$) time-window.

Estimates for the five major sources of background for the new data
sample are improved with respect to Ref.~\cite{DBPRL2014}.  The
background produced by the three Am-C neutron sources inside the
automated calibration units contributed significantly to the total
systematic uncertainty of the correlated backgrounds in the 6-AD
period. Because of this, two of the three Am-C sources in each AD in
EH3 were removed during the 2012 summer installation period. As a
result, the average correlated Am-C background rate in the far hall
decreased by a factor of 4 in the 8-AD period. As in previous
publications~\cite{DB,DBPRL2014}, this rate was determined by
monitoring the single neutron production rate from the Am-C
sources. Removal of these Am-C sources had negligible consequences for
our calibration.

Energetic, or {\it fast}, neutrons of cosmogenic origin produced a
correlated background for this study. Relaxing the prompt-energy
selection to (0.7-100)~MeV revealed the fast-neutron background
spectrum above 12~MeV.  Previously we deduced the rate and spectrum of
this background using a linear extrapolation into the IBD prompt
signal region.  Here we used a background-enhanced dataset to improve
the estimate.  We found 6043 fast neutron candidates with prompt
energy from 0.7 to 100~MeV in the 200~$\mu$s following cosmogenic
signals only detected by the OWS or RPC.  The energy spectrum of these
veto-tagged signals was consistent with the spectrum of IBD-like
candidate signals above 12~MeV, and was used to estimate the rate and
energy spectrum for the fast neutron background from 0.7 to 12~MeV.
The systematic uncertainty was estimated from the difference between
this new analysis and the extrapolation method previously employed,
and was determined to be half of the estimate reported in
Ref.~\cite{CPC}.

The methods used in Refs.~\cite{DB,CPC} to estimate the backgrounds
from the uncorrelated prompt-delayed pairs ({\it i.e.} accidentals),
the correlated $\beta$-$n$ decays from cosmogenic $^9$Li and $^{8}$He,
and the $^{13}$C($\alpha$,n)$^{16}$O reaction, were extended to the
current 6+8 AD data sample. The decrease in the single-neutron rate
from the Am-C sources reduced the average rate of accidentals in the
far hall by a factor of 2.7. As a result, the total backgrounds amount
to about 3\% (2\%) of the IBD candidate sample in the far (near)
hall(s).  The systematic uncertainties in the
$^{13}$C($\alpha$,n)$^{16}$O cross section and in the transportation
of the $\alpha$ particles were reassessed through a comparison of
experimental results and simulation packages,
respectively~\cite{AlphaN}.  The estimation of $^9$Li/$^{8}$He now
dominated the background uncertainty in both the near and far
halls. The estimated signal and background rates, as well as the
efficiencies of the muon veto, $\epsilon_{\mu}$, and multiplicity
selection, $\epsilon_m$, are summarized in Table~\ref{tab:ibd}.

\begin{table*}[!htb]
  \begin{minipage}[c]{\textwidth}
  \resizebox{\textwidth}{!}{
\begin{tabular}{c|cc|cc|cccc} \hline
\hline
  & \multicolumn{2}{c|}{EH1}&\multicolumn{2}{c|}{EH2}&\multicolumn{4}{c}{EH3} \\
  & AD1  & AD2  & AD3 & AD8 & AD4 & AD5 & AD6 & AD7 \\
\hline
IBD candidates & 304459	&309354 & 287098 & 190046 & 40956 & 41203 & 40677 & 27419 \\
DAQ live time(days) & 565.436 & 565.436 & 568.03 & 378.407 & 562.451 & 562.451 & 562.451 & 372.685 \\
$\varepsilon_{\mu}$ & 0.8248 & 0.8218	&0.8575	 &0.8577	 &0.9811	&0.9811	&0.9808	&0.9811 \\
$\varepsilon_{m}$ & 0.9744	&0.9748	&0.9758	&0.9756	 &0.9756	 &0.9754	&0.9751	&0.9758 \\
Accidentals(per day) & $8.92\pm0.09$ & $8.94\pm0.09$ & $6.76\pm0.07$ & $6.86\pm0.07$ & $1.70\pm0.02$ & $1.59\pm0.02$ & $1.57\pm0.02$ & $1.26\pm0.01$ \\
Fast neutron(per AD per day) & \multicolumn{2}{c|}{$0.78\pm0.12$} & \multicolumn{2}{c|}{$0.54\pm0.19$} & \multicolumn{4}{c}{$0.05\pm0.01$} \\
$^9$Li/$^8$He(per AD per day) & \multicolumn{2}{c|}{$2.8\pm1.5$} & \multicolumn{2}{c|}{$1.7\pm0.9$} & \multicolumn{4}{c}{$0.27\pm0.14$} \\
Am-C correlated 6-AD(per day) & $0.27\pm0.12$ & $0.25\pm0.11$ & $0.27\pm0.12$ & & $0.22\pm0.10$ & $0.21\pm0.10$ & $0.21\pm0.09$  \\
Am-C correlated 8-AD(per day) & $0.20\pm0.09$ & $0.21\pm0.10$ & $0.18\pm0.08$ & $0.22\pm0.10$ & $0.06\pm0.03$ & $0.04\pm0.02$ & $0.04\pm0.02$ & $0.07\pm0.03$ \\
$^{13}$C($\alpha$, n)$^{16}$O(per day) & $0.08\pm0.04$ & $0.07\pm0.04$ & $0.05\pm0.03$ & $0.07\pm0.04$ & $0.05\pm0.03$ & $0.05\pm0.03$ & $0.05\pm0.03$ & $0.05\pm0.03$ \\ \hline
IBD rate(per day) & $657.18\pm1.94$ & $670.14\pm1.95$ & $594.78\pm1.46$	& $590.81\pm1.66$ & $73.90\pm0.41$ & $74.49\pm0.41$ & $73.58\pm0.40$	& $75.15\pm0.49$ \\ 
\hline
\end{tabular}}
  \end{minipage}
  \caption{Summary of signal and backgrounds. Rates are corrected for
    the muon veto and multiplicity selection efficiencies
    $\varepsilon_{\mu}\cdot\varepsilon_{m}$.  The measured ratio of
    the IBD rates in AD1 and AD2 (AD3 and AD8 in the 8-AD period) was
    0.981$\pm$0.004 (1.019$\pm$0.004) while the expected ratio was
    0.982 (1.012).
\label{tab:ibd}}
\end{table*}
A detailed treatment of the absolute and relative efficiencies using
the first six ADs was reported in Refs.~\cite{DBNIM, CPC}. The
uncertainties of the absolute efficiencies are correlated among the
ADs and thus play a negligible role in the relative measurement of
$\overline{\nu}_{e}$ disappearance.  The performance of the two new
ADs was found to be consistent with the other detectors.  Estimates of
two prominent uncorrelated uncertainties, the delayed-energy selection
efficiency and the fraction of neutrons captured on Gd, were confirmed
for all eight ADs using improved energy reconstruction and increased
statistics.

Oscillation was measured using the $L/E$-dependent disappearance of
$\overline{\nu}_e$, as given by the survival probability
\begin{eqnarray}\label{eq:psurv}
P = 1 &-& \cos^4\theta_{13}\sin^2 2\theta_{12}\sin^2\frac{1.267\Delta
  m^2_{21}L}{E} \hfill \nonumber \\ &-& \sin^2 2\theta_{13}\sin^2
\frac{1.267\Delta m^2_{ee}L}{E}.
\end{eqnarray}
Here $E$ is the energy in MeV of the \nuebar, $L$ is the distance in
meters from its production point, $\theta_{12}$ is the solar mixing
angle, and ${\Delta}m^2_{21}=m^{2}_{2}-m^{2}_{1}$ is the mass-squared
difference of the first two neutrino mass eigenstates in eV$^2$.

Recent precise measurements of the IBD positron energy spectrum
disagree with models of reactor $\overline{\nu}_e$
emission~\cite{Abe:2014bwa,zhongwl:ihep2014,reno:nu2014,Dwyer:2014eka}.
The characteristics of the signals in this energy range are consistent with reactor antineutrino emission, and disfavor background or detector response as possible origins for the discrepancy. A separate manuscript, in preparation, will present the evidence in detail and provide the necessary data to allow detailed comparison of our measurement with existing and future models. Given these discrepancies between measurements and models,
here we present a technique for predicting the signal in the far hall
based on measurements obtained in the near halls, with minimal
dependence on models of the reactor antineutrinos.  In our previous
measurements~\cite{DBPRL2014}, model-dependence was limited by
allowing variation of the predicted $\overline{\nu}_e$ flux within
model uncertainties, while the technique here provides an explicit
demonstration of the negligible model dependence.  A $\chi^2$ was
defined as
\begin{equation}
  \label{eq:chi2}
  \chi^{2} = \sum_{i,j}(N_j^{\mathrm{f}} - w_j \cdot N_j^{\mathrm{n}}) (V^{-1})_{ij} (N_i^{\mathrm{f}} - w_i \cdot N_i^{\mathrm{n}}),
\end{equation}
where $N_i$ is the observed number of events after background
subtraction in the $i$-th bin of reconstructed positron energy
$E^{\mathrm{rec}}$\@. The superscript $f~(n)$ denotes a far~(near)
detector. The symbol $V$ represents a covariance matrix that includes
known systematic and statistical uncertainties.  The quantity $w_i$ is
a weight that accounts for the differences between near and far
measurements. For the case of a single reactor, the weight $w_i$ can
be simply calculated from the ratios of detector mass, distance to the
reactor, efficiency, and antineutrino oscillation probability, as
given by the relation:
\begin{equation}\label{eq:simplecase}
{w}_i^{\mathrm{SR}} = \frac{N^{\mathrm{f}}_i}{N^{\mathrm{n}}_i} =
\left( \frac{T^{\mathrm{f}}}{T^{\mathrm{n}}} \right) \left(
\frac{\epsilon^{\mathrm{f}}}{\epsilon^{\mathrm{n}}} \right) \left( \frac{L^{\mathrm{n}}}{L^{\mathrm{f}}}
\right)^2
\left( \frac{ P^{\mathrm{f}}_i}{ P^{\mathrm{n}}_i} \right)
  {\left(\frac{\phi}{\phi}\right)}.
\end{equation}
Here $T$ is the number of target protons, $\epsilon$ is the
efficiency, and $L$ is the distance to the reactor for a given
detector. $P_i$ is the oscillation probability for the $i$-th
reconstructed energy bin and $\phi$ the reactor antineutrino flux
(which cancels from $w_{i}$).  With $P_i$ calculated in reconstructed
positron energy, the detector response introduces small ($<$~0.2\%
above 2~MeV) calculable deviations from Eq.~\ref{eq:psurv}.

For multiple reactor cores, the weight $w_i$ was modified:
\begin{equation}
\label{eq:dybwi}
w_i = \frac{N^{\mathrm{f}}_i}{N^{\mathrm{n}}_i} = \left(
\frac{T^{\mathrm{f}}}{T^{\mathrm{n}}} \right) \left(
\frac{\epsilon^{\mathrm{f}}}{\epsilon^{\mathrm{n}}} \right)
\sum_j \mathcal{P}(E^{\mathrm{true}}_j|E^{\mathrm{rec}}_i) r_j.
\end{equation}
The probability distribution
$\mathcal{P}(E^{\mathrm{true}}_j|E^{\mathrm{rec}}_i)$ accounts for the
energy transfer from the \nuebar\ to the $e^+$ and imperfections in
the detector energy response (loss in non-active elements,
non-linearity, and resolution).  The extrapolation factor $r_j$ was
calculated as
\begin{equation}
  \label{eq:far-near-extrapolation}
  r_{j} = \frac{\sum^{\mathrm{cores}}_{k}P(E^{\mathrm{true}}_j,L^{\mathrm{f}}_k)\phi_{jk}/(L^{\mathrm{f}}_k)^2}
  {\sum^{\mathrm{cores}}_{k}P(E^{\mathrm{true}}_j,L^{\mathrm{n}}_k)\phi_{jk}/(L^{\mathrm{n}}_k)^2},
\end{equation}
where $P$ is given by Eq.~\ref{eq:psurv}, $L^{f(n)}_k$ is the distance
between a far~(near) detector and core $k$, and $\phi_{jk}$ is the
predicted antineutrino flux from core $k$ for the $j$-th true energy
bin. In the single-reactor core case, the antineutrino flux $\phi$
cancels in the expression for $r_j$ and Eq.~\ref{eq:dybwi} reduces to
Eq.~\ref{eq:simplecase}. Although the cancellation is not exact for
multiple cores, the impact of the uncertainty in reactor antineutrino
flux was found to be $\leq$~0.1\%.

The covariance matrix element $V_{ij}$ was the sum of a statistical
term, calculated analytically, and a systematic term determined by
Monte-Carlo calculation using
\begin{equation}\label{eq:covmat}
V_{ij} = \frac{1}{N} \sum^{N}
\left( S^{\mathrm{f}}_i - w_{i} \cdot S^{\mathrm{n}}_i  \right)
 \left( S^{\mathrm{f}}_j - w_{j} \cdot S^{\mathrm{n}}_j  \right).
\end{equation}
Here, $N$ is the number of simulated experiments generated with energy
spectra $S$, including systematic variations of detector response,
{\nuebar} flux, and background.  The choice of reactor antineutrino
model~\cite{Dwyer:2014eka,Schrek1,Schrek2,Schrek3,Vogel1,HuberAnomaly,Mueller} in
calculating the covariance had negligible ($<$0.2\%) impact on the determination
of the oscillation parameters.

Without loss of sensitivity, we summed the IBD signal candidates of
the ADs within the same hall, accounting for small differences of
target mass, detection efficiency, background and baseline. We
considered the 6-AD and 8-AD periods separately in order to properly
handle correlations in reactor antineutrino flux, detector exposure,
and background. This means that $i$ and $j$ in the above equations ran
over the 37 reconstructed energy bins for the two near/far
combinations and for the two periods considered
($37\times2\times2=148$).  More details of this method are described
in Ref.~\cite{BerkeleyFitter}.


Using this method, we found $\sin^22\theta_{13}=0.084\pm0.005$ and
$|\Delta m^2_{\mathrm{ee}}|=(2.42\pm0.11)\times10^{-3}~\mathrm{eV}^2$,
with $\chi^2/\mathrm{NDF}=134.6/146$ (see the Supplemental Material~\footnote{See Supplemental Material at [URL] for a table of $\chi^2-\chi^2_{\rm min}$
as a function of ($\sin^{2}2\theta_{13}$, $|\Delta m^2_{\mathrm{ee}}|$)}). While we use $\sin^2
2\theta_{12}=0.857\pm0.024$ and $\Delta
m^2_{21}=(7.50\pm0.20)\times10^{-5}~\mathrm{eV}^2$ from
Ref.~\cite{PDG}, our result was largely independent of these
values. Consistent results were obtained when our previous methods
\cite{DB,DBPRL2014} were applied to this larger dataset. Under the
normal (inverted) hierarchy assumption, $|\Delta m^2_{\mathrm{ee}}|$
yields $\Delta
m^2_{32}=(2.37\pm0.11)\times10^{-3}~\mathrm{eV}^2$~($\Delta
m^2_{32}=-(2.47\pm0.11)\times10^{-3}~\mathrm{eV}^2$).  This result was
consistent with and of comparable precision to measurements obtained
from accelerator $\nu_\mu$ and $\bar{\nu}_\mu$
disappearance~\cite{Adamson:2014vgd, Abe:2014ugx}.  Using only the
relative rates between the detectors and $\Delta m^2_{32}$ from
Ref.~\cite{Adamson:2014vgd} we found
$\sin^22\theta_{13}=0.085\pm0.006$, with $\chi^2/\mathrm{NDF}=1.37/3$.

\begin{figure}[htb]
\centering
\includegraphics[width=0.95\columnwidth]{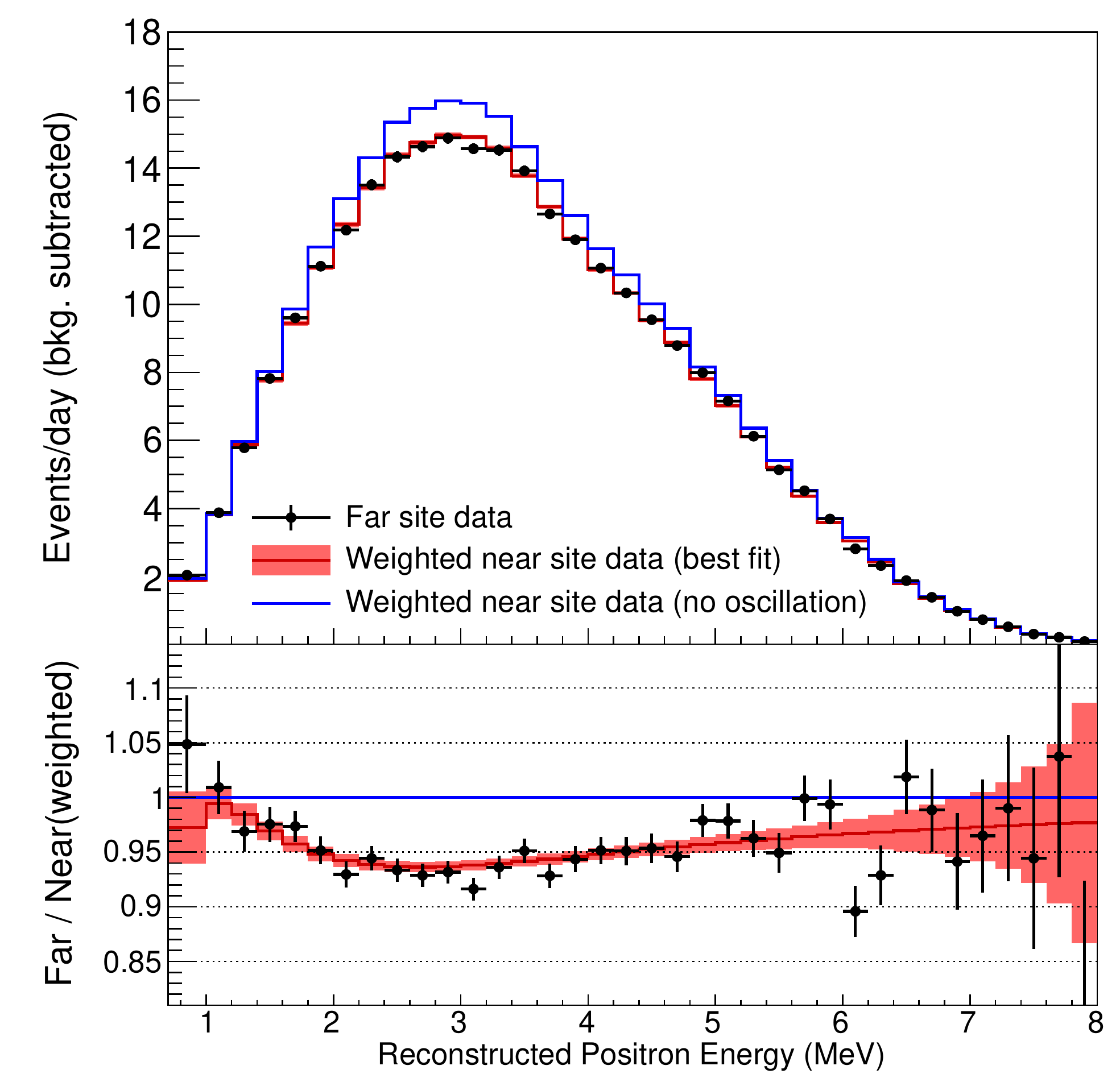}
\caption{Upper: Background-subtracted reconstructed positron energy
  spectrum observed in the far site (black points), as well as the
  expectation derived from the near sites excluding (blue line) or
  including (red line) our best estimate of oscillation. The spectra
  were efficiency-corrected and normalized to one day of
  livetime. Lower: Ratio of the spectra to the no-oscillation
  case. The error bars show the statistical uncertainty of the far
  site data. The shaded area includes the systematic and statistical
  uncertainties from the near site measurements.
\label{fig:spectracomp}}
\end{figure}

\begin{figure}[htb]
\centering
\includegraphics[width=0.95\columnwidth]{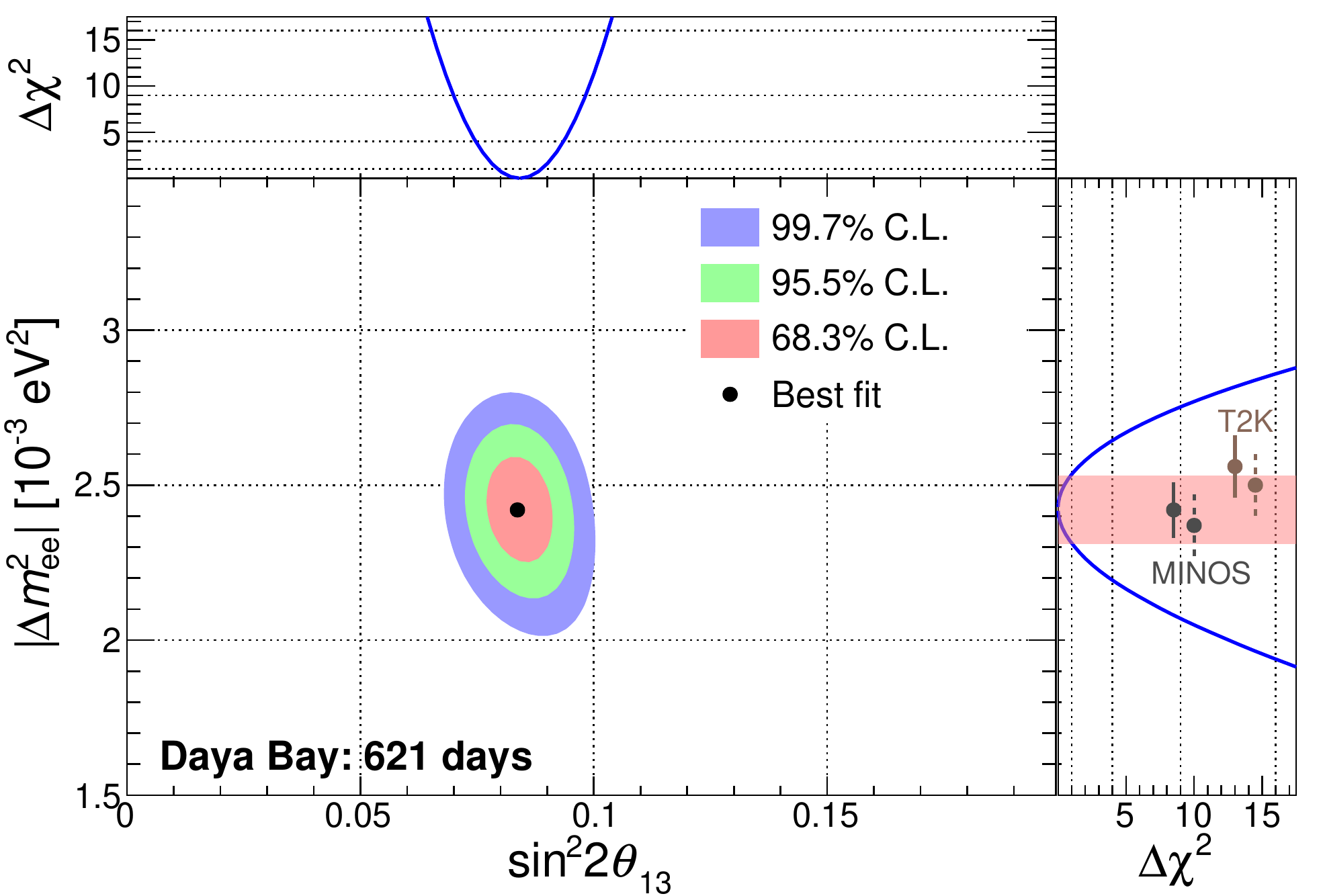}
\caption{Regions in the $|\Delta
  m^{2}_{\mathrm{ee}}|$-$\sin^{2}2\theta_{13}$ plane allowed at the
  $68.3\%$, $95.5\%$ and $99.7\%$ confidence levels by the near-far
  comparison of $\overline{\nu}_e$ rate and energy spectra. The best
  estimates were $\sin^22\theta_{13}=0.084\pm0.005$ and $|\Delta
  m^2_{\mathrm{ee}}|=(2.42\pm0.11)\times10^{-3}~\mathrm{eV}^2$ (black
  point). The adjoining panels show the dependence of $\Delta
  \chi^{2}$ on $\sin^{2}2\theta_{13}$ (top) and $|\Delta
  m^2_{\mathrm{ee}}|$ (right).  The $|\Delta m^2_{\mathrm{ee}}|$
  allowed region (shaded band, $68.3\%$ C.L.) was consistent with
  measurements of $|\Delta m^2_{32}|$ using muon disappearance by the
  MINOS~\cite{Adamson:2014vgd} and T2K~\cite{Abe:2014ugx} experiments,
  converted to $|\Delta m^2_{\mathrm{ee}}|$ assuming the normal
  (solid) and inverted (dashed) mass hierarchy.}
\label{fig:contours}
\end{figure}

\begin{figure}[htb]
\centering
\includegraphics[width=\columnwidth]{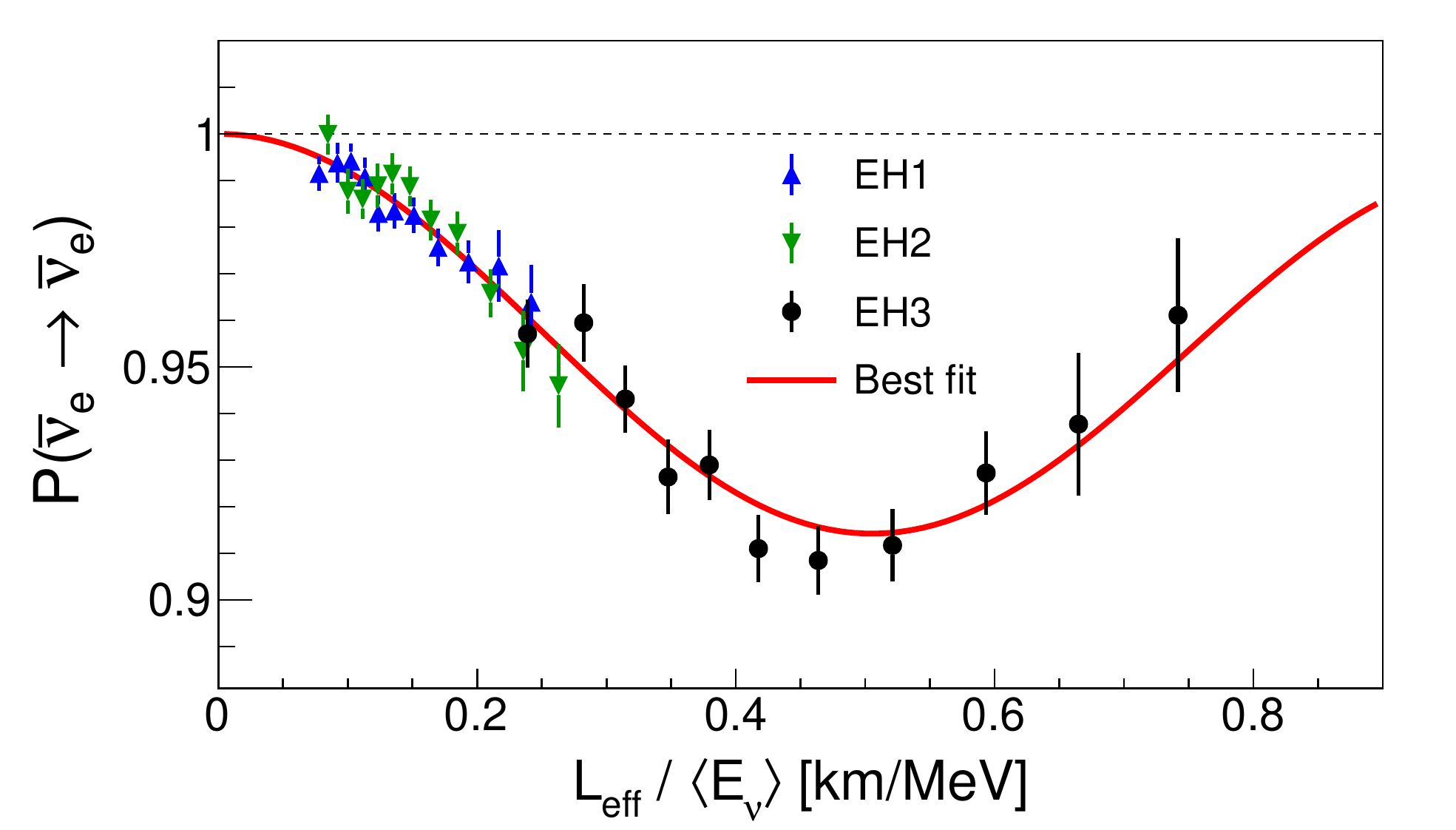}
\caption{Electron antineutrino survival probability versus effective
  propagation distance $L_{\mathrm{eff}}$ divided by the average
  antineutrino energy $\langle E_{\nu} \rangle$. The data points
  represent the ratios of the observed antineutrino spectra to the
  expectation assuming no oscillation. The solid line represents the
  expectation using the best estimates of $\sin^22\theta_{13}$ and
  $|\Delta m^2_{ee}|$. The error bars are statistical only.  $\langle
  E_{\nu} \rangle$ was calculated for each bin using the estimated
  detector response, and $L_{\mathrm{eff}}$ was obtained by equating
  the actual flux to an effective antineutrino flux using a single
  baseline.}
\label{fig:loe}
\end{figure}

The reconstructed positron energy spectrum observed in the far site is
compared in Fig.~\ref{fig:spectracomp} with the expectation based on
the near-site measurements.  The $68.3\%$, $95.5\%$ and $99.7\%$
C.L. allowed regions in the $|\Delta
m^{2}_{\mathrm{ee}}|$-$\sin^{2}2\theta_{13}$ plane are shown in
Fig.~\ref{fig:contours}.  The spectral shape from all experimental
halls is compared in Fig.~\ref{fig:loe} to the electron antineutrino
survival probability assuming our best estimates of the oscillation
parameters.  The total uncertainties of both $\sin^{2}2\theta_{13}$
and $|\Delta m^{2}_{\mathrm{ee}}|$ are dominated by statistics. The
most significant systematic uncertainties for $\sin^22\theta_{13}$ are
due to the relative detector efficiency, reactor power, relative
energy scale and $^{9}$Li/$^{8}$He background.  The systematic
uncertainty in $|\Delta m^2_{\mathrm{ee}}|$ is dominated by
uncertainty in the relative energy scale.


In summary, enhanced measurements of $\sin^22\theta_{13}$ and $|\Delta
m^2_{ee}|$ have been obtained by studying the energy-dependent
disappearance of the electron antineutrino interactions recorded in a
6.9$\times$10$^5$ GW$_{\rm th}$-ton-days exposure. Improvements in
calibration, background estimation, as well as increased statistics
allow this study to provide the most precise estimates to date of the
neutrino mass and mixing parameters $|\Delta m^2_{ee}|$ and
$\sin^22\theta_{13}$.

Daya Bay is supported in part by the Ministry of Science and
Technology of China, the U.S. Department of Energy, the Chinese
Academy of Sciences, the CAS Center for Excellence in Particle
Physics, the National Natural Science Foundation of China, the
Guangdong provincial government, the Shenzhen municipal government,
the China General Nuclear Power Group, Key Laboratory of Particle and
Radiation Imaging (Tsinghua University), the Ministry of Education,
Key Laboratory of Particle Physics and Particle Irradiation (Shandong
University), the Ministry of Education, Shanghai Laboratory for
Particle Physics and Cosmology, the Research Grants Council of the
Hong Kong Special Administrative Region of China, the University
Development Fund of The University of Hong Kong, the MOE program for
Research of Excellence at National Taiwan University, National
Chiao-Tung University, and NSC fund support from Taiwan, the
U.S. National Science Foundation, the Alfred~P.~Sloan Foundation, the
Ministry of Education, Youth, and Sports of the Czech Republic, the
Joint Institute of Nuclear Research in Dubna, Russia, the NSFC-RFBR
joint research program, the National Commission of Scientific and
Technological Research of Chile, and the Tsinghua University
Initiative Scientific Research Program. We acknowledge Yellow River
Engineering Consulting Co., Ltd., and China Railway 15th Bureau Group
Co., Ltd., for building the underground laboratory. We are grateful
for the ongoing cooperation from the China General Nuclear Power Group
and China Light and Power Company.


\appendix

\section{Appendix: Why $\Delta m^2_{ee}$ is used by Daya Bay}

This section describes the advantages of reporting the Daya Bay
  measurement of electron antineutrino disappearance in terms of an
  effective mass-squared difference ${\Delta}m^2_{ee}$, which is
  independent of the unknown ordering of neutrino masses and future
  improvements in our knowledge of the solar oscillation parameters.

\subsection{Introduction}

In the three-flavor framework, the survival probability of electron antineutrino
is given by
\begin{align}
  \label{eq:exact}
 & P(\overline{\nu}_e \to \overline{\nu}_e ) \nonumber \\
 & ~~=  1   - \cos^4\theta_{13}\sin^22\theta_{12}\sin^2\Delta_{21}  \nonumber \\
 &~~~~~ - \sin^22\theta_{13}
  (\cos^2\theta_{12}\sin^2 \Delta_{31} + \sin^2\theta_{12}\sin^2\Delta_{32}),
\end{align}
where $\Delta_x = \Delta m^2_x \frac{L}{4E}$. The three
mass-squared differences are subject to the constraint $|\Delta
m^2_{31}| = |\Delta m^2_{32}| \pm |\Delta m^2_{21}|$ where
``$+$''(``$-$'') is for the normal(inverted) mass ordering (or
  hierarchy).  Therefore, determination of $\Delta m^2_{32}$
(or $\Delta m^2_{31}$) depends on knowledge of the mass
ordering and solar oscillation parameters.

The Daya Bay experiment reports a precise measurement of the
  effective mass splitting $\Delta m^2_{ee}$, which is independent of
  our knowledge of the ordering and solar parameters.  In this
  approach, we approximate the survival probability using
\begin{align}
  \label{eq:approx}
  P(\overline{\nu}_e \to \overline{\nu}_e )
  \simeq&  1
  - \cos^4\theta_{13}\sin^22\theta_{12}\sin^2\Delta_{21} \nonumber \\
&- \sin^22\theta_{13}   \sin^2 \Delta_{ee}.
\end{align}
Despite the advantage of using $\Delta m^2_{ee}$ for the
measurement, it has the disadvantage of not being a fundamental
parameter. Therefore, we must determine a relation between
  $\Delta m^2_{ee}$ and $\Delta m^2_{32}$ given knowledge of the mass
  ordering and solar oscillation parameters.

In the following sections, we are going to address the following two questions:
\begin{itemize}
\item Is Eq.~\ref{eq:approx} good enough at the current experimental precision?
\item How can we estimate the value of $\Delta m^2_{32}$ once the
  value of $\Delta m^2_{ee}$ is obtained?
\end{itemize}

\subsection{Mathematical derivation}
Using the relation  $|\Delta m^2_{31}| = |\Delta m^2_{32}| \pm
|\Delta m^2_{21}|$, Eq.~\ref{eq:exact} can be written as,
\begin{align}
  \label{eq:1}
  P(\overline{\nu}_e \to \overline{\nu}_e )
= &~1 - 2s^2_{13}c^2_{13}  \nonumber\\
&+ 2s^2_{13}c^2_{13}\sqrt{1-4s^2_{12}c^2_{12}\sin^2\Delta_{21}}
\cos (2\Delta_{32} \pm \phi) \nonumber\\
& - 4 c^4_{13}s^2_{12}c^2_{12}\sin^2\Delta_{21},
\end{align}
where $s_{x}  = \sin \theta_{x}$, $c_{x}  = \cos \theta_{x}$,
 and $\phi = \arctan\left( \frac{\sin 2\Delta_{21}}
{\cos 2\Delta_{21} + \tan^2 \theta_{12}}\right)$.
The last term of the above formula is the so-called ``solar term''
that governs the reactor antineutrino oscillation at O(100)~km.
For the L/E range covered by Daya Bay, $4s^2_{12}c^2_{12}\sin^2\Delta_{21} \ll 1$.
Thus, Eq.~\ref{eq:1} can be approximated as,
\begin{align}
  \label{eq:3}
&P(\overline{\nu}_e \to \overline{\nu}_e ) \nonumber \\
& \simeq 1 - 4 s^2_{13}c^2_{13} \left[\frac{1-\cos(2\Delta_{32} \pm \phi)}{2}\right]
- (solar~term)\nonumber \\
& = 1 - \sin^22\theta_{13} \sin^2(\Delta_{32}\pm \phi/2) - (solar~term).
\end{align}

By comparing Eq.~\ref{eq:3} with Eq.~\ref{eq:approx}, we obtain
the expression relating $\Delta m^2_{ee}$ to $\Delta m^2_{32}$ (or $\Delta m^2_{31}$)
\begin{align}
  \label{eq:4}
  |\Delta m^2_{ee}| &= |\Delta m^2_{32}| \pm \Delta m^2_\phi/2\\
  &= |\Delta  m^2_{31}| \mp (|\Delta m^2_{21}| - \Delta m^2_\phi/2),
  \label{eq:4a}
\end{align}
where $\Delta m^2_\phi = \phi \times \frac{4E}{L}$.

\subsection{Numerical evaluation}

\begin{figure}[hbtp]
  \centering
  \includegraphics[width=0.45\textwidth]{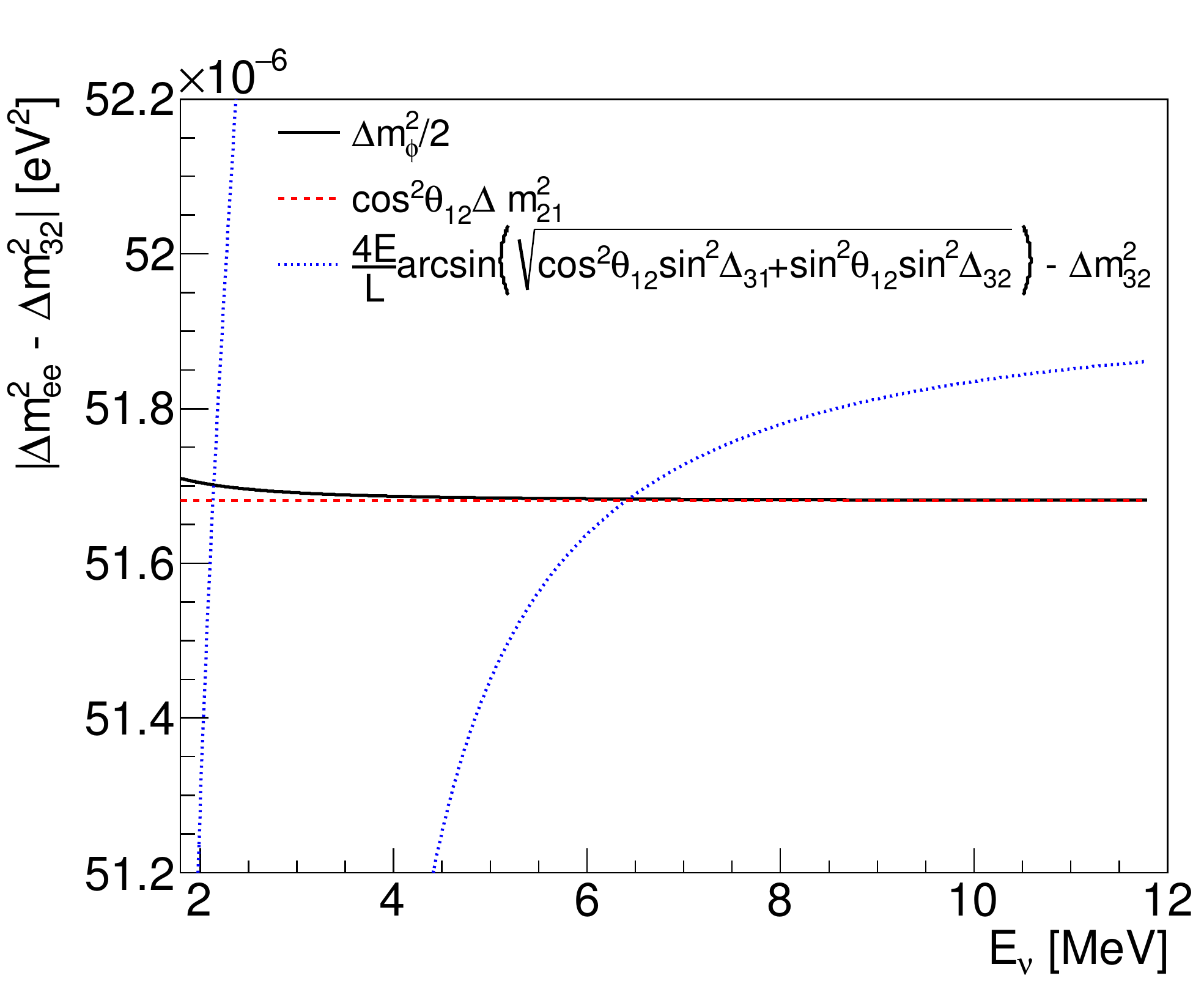}
  \caption{Values of $\Delta m^2_\phi/2 = |\Delta m^2_{ee} -
    \Delta m^2_{32}|$ (black solid line) at L~=~1.6~km as a function of the neutrino
  energy, with $\Delta m^2_{21} = 7.50 \times 10^{-5}~\textrm{eV}^2$
  and $\sin^22\theta_{12} = 0.857$~\cite{PDG}.
  For comparison, calculations based on other definitions of
      $\Delta m^2_{ee}$,
      $\Delta m^2_{ee} = \cos^2 \theta_{12}\Delta m^2_{31} + \sin^2
      \theta_{12}\Delta m^2_{32}$ (red dashed line) and
      $\Delta m^2_{ee} = \frac{4E}{L}\arcsin \left[\sqrt{\cos^2\theta_{12}\sin^2 \Delta_{31}
          + \sin^2\theta_{12}\sin^2 \Delta_{32} }\right]$ (blue dotted line)
      are also shown.
  }
\label{fig:dm2_phi}
\end{figure}

By definition, $\Delta m^2_\phi$ is a function of
L/E. Using the current values of
$\Delta m^2_{21} = 7.50 \times 10^{-5}~\textrm{eV}^2$
and $\sin^22\theta_{12} = 0.857$~\cite{PDG},
Fig.~\ref{fig:dm2_phi} shows the value of
$\Delta m^2_\phi / 2$ as a function of energy for L~=~1.6~km.
We find that $\Delta m^2_\phi / 2 \simeq 5.17\times10^{-5}~\textrm{eV}^2$
is essentially a constant in our L/E region, and numerically identical to
cos$^2\theta_{12}\Delta m^2_{21}$.
Thus, this definition of $\Delta m^2_{ee}$ is similar to the definition
introduced in Ref.~\cite{Nunokawa:2005nx}:
\begin{align}
  \label{eq:6}
  \Delta m^2_{\textrm{eff}}|_e &= \cos^2 \theta_{12}|\Delta m^2_{31}| +
                                 \sin^2 \theta_{12}|\Delta m^2_{32}| \\
  &= |\Delta m^2_{32}| \pm \cos^2 \theta_{12} \Delta m^2_{21}.
\end{align}

\begin{figure}[hbtp]
  \centering
  \includegraphics[width=0.45\textwidth]{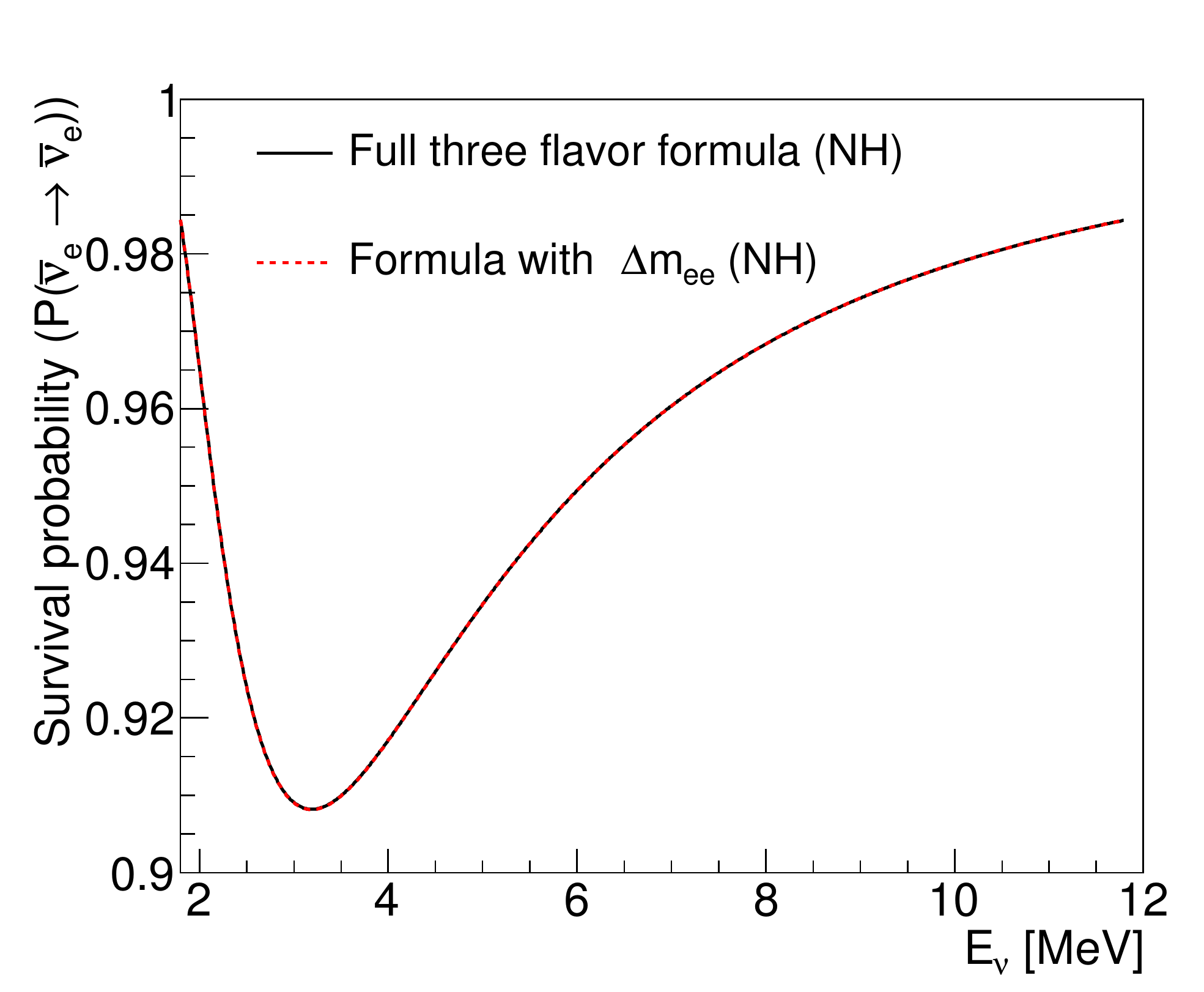}
  \includegraphics[width=0.45\textwidth]{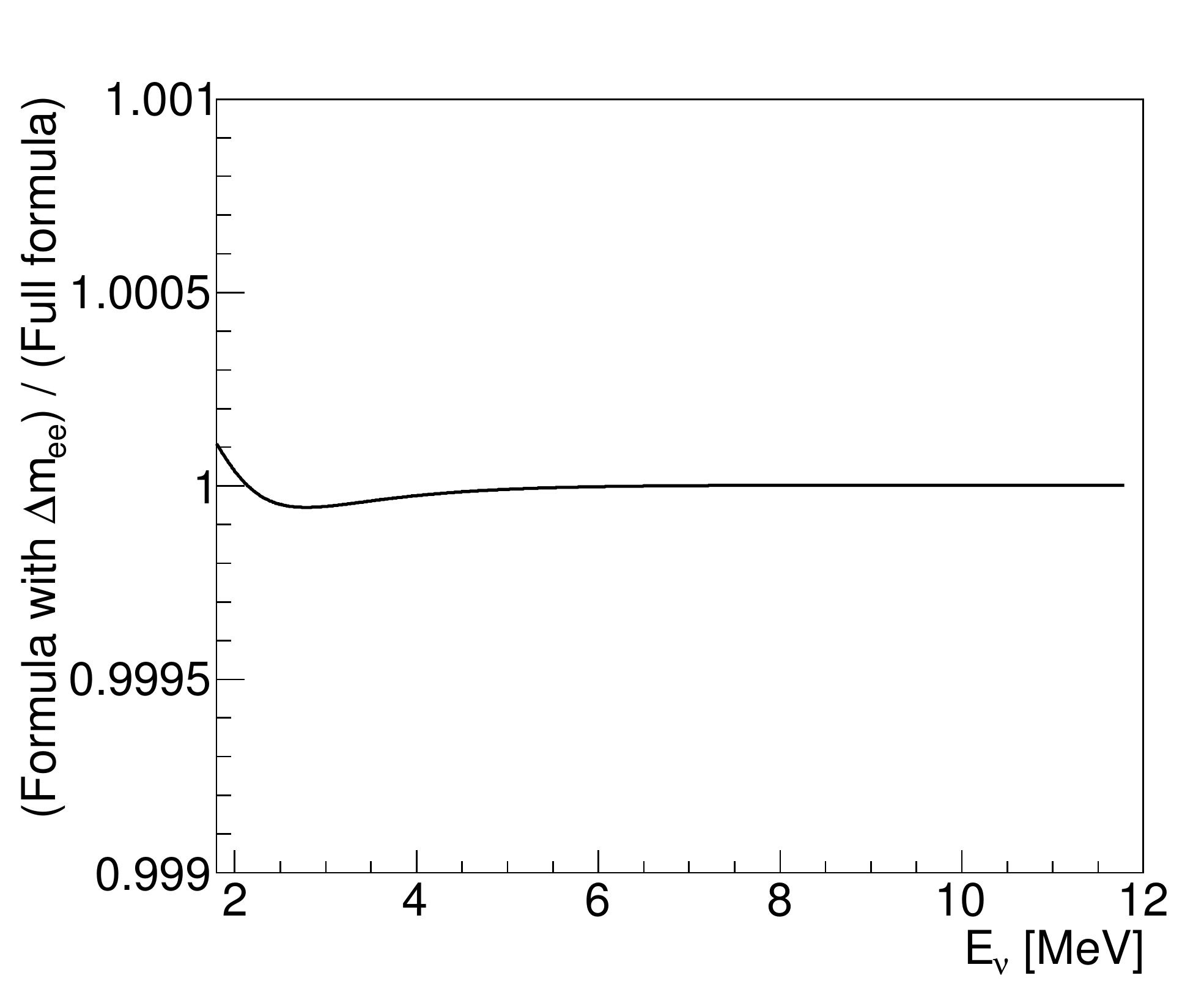}
  \caption{Comparison of the survival probability
    at L~=~1.6~km between the
    approximated formula with
    $\Delta m^2_{ee} = \Delta m^2_{32} +
    5.17\times10^{-5}~\textrm{eV}^2$ and the exact
    three-flavor formula (Eq.~\ref{eq:exact}).
    The oscillation parameters used in this comparison are $\sin^2 2\theta_{13} = 0.09$
    and $\Delta m^2_{32} = 2.44 \times 10^{-3}~\textrm{eV}^2$ under the
    normal mass hierarchy assumption.
    The top panel shows the survival probabilities calculated with the two formulae, and
    the bottom panel shows the ratio of the two.}
  \label{fig:osc_prob}
\end{figure}

Figure~\ref{fig:osc_prob} is a comparison of the approximated formula with
$\Delta m^2_\phi / 2 = 5.17\times10^{-5}~\textrm{eV}^2$,
\begin{align}
  \label{eq:5}
  P_{ee} \simeq&~1 - \sin^22\theta_{13} \sin^2\left[(\Delta m^2_{32} +
  5.17\times10^{-5}~\textrm{eV}^2)
  \frac{L}{4E}\right]\nonumber \\
&- (solar~term),
\end{align}
to the three-flavor formula, Eq.~\ref{eq:exact}.
In this comparison, L~=~1.6~km, $\sin^2 2\theta_{13} = 0.09$,
$\Delta m^2_{32} = 2.44 \times 10^{-3}~\textrm{eV}^2$, and
normal mass hierarchy are the inputs.
The agreement between the two, better than $10^{-4}$, is excellent and exceeds the
achievable experimental precision.

This study demonstrates that, once we obtain the value of $|\Delta
m^2_{ee}|$ using Eq.~\ref{eq:approx},
we can reliably deduce the values of $|\Delta m^2_{32}|$ and $|\Delta
m^2_{31}|$ using Eqs.~\ref{eq:4} and~\ref{eq:4a} with
\begin{equation}
 \label{eq:7}
\Delta m^2_\phi /2
\simeq \cos^2\theta_{12}\Delta m^2_{21}.
\end{equation}
Using the current values of $\theta_{12}$ and ${\Delta}m^2_{21}$,
$\Delta m^2_\phi /2 \simeq 5.17 \times 10^{-5}{\rm eV}^2$, and
($|\Delta m^2_{21}| - \Delta m^2_\phi /2) \simeq 2.33 \times 10^{-5}\,{\rm eV}^2$.

It is important to point out that the exact solution of
$\sin^2(\Delta m^2_{ee}
  \frac{L}{4E})
= \cos^2\theta_{12}\sin^2(\Delta m^2_{31}  \frac{L}{4E})
  + \sin^2\theta_{12}\sin^2(\Delta m^2_{32} \frac{L}{4E})$
was never used to extract
the value of $\Delta m^2_{32}$ or $\Delta m^2_{31}$ from the measured
$\Delta m^2_{ee}$ in Daya Bay.

\bibliographystyle{apsrev4-1}
\bibliography{DYB_8AD_osc}

\end{document}